\documentclass[acmsmall,screen]{acmart}
\usepackage{tabularx}  
\usepackage{subcaption}
\usepackage[most]{tcolorbox}
\AtBeginDocument{%
  }

\setcopyright{acmlicensed}
\copyrightyear{2026}
\acmYear{2026}

\newcommand{\RQI}{How do logging practices in agentic pull requests differ from those in human pull requests?}

\newcommand{\RQII}{How prevalent are explicit logging instructions in issue descriptions and repository agent-instruction files?}

\newcommand{\RQIII}{Is agentic logging regulated post generation, and by whom?}

\begin{document}

\title{Do AI Coding Agents Log Like Humans? An Empirical Study}

\author{Youssef Esseddiq Ouatiti}
\email{youssefesseddiq.ouatiti@queensu.ca}
\affiliation{%
  \institution{Queen's University}
  \city{Kingston}
  \country{Canada}
}

\author{Mohammed Sayagh}
\affiliation{%
  \institution{ETS - Québec University}
  \city{Montreal}
  \country{Canada}}
\email{mohammed.sayagh@etsmtl.ca}

\author{Hao Li}
\affiliation{%
  \institution{Queen's University}
  \city{Kingston}
  \country{Canada}
}
\email{hao.li@queensu.ca}

\author{Ahmed E. Hassan}
\affiliation{%
  \institution{Queen's University}
  \city{Kingston}
  \country{Canada}
}
\email{ahmed@cs.queensu.ca}
\renewcommand{\shortauthors}{Ouatiti et al.}



\begin{abstract} 
Software logging is essential for maintaining and debugging complex systems, yet it remains unclear how AI coding agents handle this non-functional requirement. While prior work characterizes human logging practices, the behaviors of AI coding agents and the efficacy of natural language instructions in governing them are unexplored. To address this gap, we conduct an empirical study of 4,550 agentic pull requests across 81 open-source repositories. We compare agent logging patterns against human baselines and analyze the impact of explicit logging instructions. We find that agents change logging less often than humans in 58.4\% of repositories, though they exhibit higher log density when they do. Furthermore, explicit logging instructions are rare (4.7\%) and ineffective, as agents fail to comply with constructive requests 67\% of the time. Finally, we observe that humans perform 72.5\% of post-generation log repairs, acting as ``silent janitors'' who fix logging and observability issues without explicit review feedback. These findings indicate a dual failure in natural language instruction (i.e., scarcity of logging instructions and low agent compliance), suggesting that deterministic guardrails might be necessary to ensure consistent logging practices.
\end{abstract}

\maketitle

\section{Introduction}
Large Language Models (LLMs) are transforming software engineering by enabling AI coding agents to generate and submit code changes. Unlike simple code completion tools, these agents interpret high-level goals, plan tasks, and execute pull requests (PRs) with minimal human intervention~\cite{Yang:2024,Watanabe:2025,Zhang:2024}. However, as agents take on more responsibility, they must adhere not only to functional requirements (i.e., passing tests) but also to non-functional requirements (NFRs) such as observability.

Observability is a critical NFR for diagnosing failures and monitoring system health~\cite{Li:2020,Shang:2015,Yuan:2012}, and it is primarily realized through logging. Yet, in traditional development, logging practices are often informal and learned through experience or tribal knowledge~\cite{Rong:2023}. Additionally, developers must balance the trade-off between providing sufficient context and avoiding excessive noise~\cite{Yuan:2012,Li:2020}. For instance, a lack of logging leads to limited runtime information and a reduced ability to diagnose issues~\cite{Yuan:2012a}. Logging too much, however, can cause system overhead and make logs noisy and difficult to analyze~\cite{Yuan:2014}. It remains unknown whether AI agents can navigate these trade-offs or whether they simply replicate insecure or overly verbose patterns present in their training data and the repository environments in which they operate.

This uncertainty presents a significant gap in our understanding of agentic logging. While recent studies have examined the functional correctness and acceptance rates of agentic PRs~\cite{Horikawa:2025, Watanabe:2024}, the observability gap remains unaddressed. For instance, it is unknown whether agents mimic human logging habits or whether developers effectively instruct agents to maintain logging and observability standards in the first place. Without this knowledge, practitioners risk integrating agents that produce opaque and unmaintainable code.

To address this gap, we conduct an empirical study of logging practices in agent-generated code. We leverage the AIDev dataset~\cite{Li:2025} to analyze 4,550 agentic PRs and 3,276 human PRs across 81 well-maintained, mature, and popular repositories. We combine quantitative metrics with qualitative analysis of instructions and review comments to characterize the entire lifecycle of agentic logging. Our study addresses the following research questions~(RQs):

\textbf{RQ1. \RQI} We observe that agents change logging less often than humans in 58.4\% of the studied repositories. However, in repositories where both agents and humans add logs, agents introduce 30\% more logs per 1,000 lines of code. While these agents successfully mimic human error-logging patterns, they are less consistent in matching human use of informational context (e.g., \texttt{INFO} level statements).

\textbf{RQ2. \RQII} We find that logging instructions are rare (4.7\%) and largely ineffective. For instance, agents fail to comply with logging requests 67\% of the time, regardless of how specific those logging instructions are.

\textbf{RQ3. \RQIII} We observe a hidden maintenance burden, as humans perform 72.5\% of post-generation logging repairs. This regulation is mostly implicit, with humans fixing logging and observability issues in subsequent commits rather than requesting changes during code review.

This paper contributes the first comparative analysis of logging practices between human and agentic contributors across mature and popular software repositories. Furthermore, it highlights the logging instruction gap and evaluates the compliance gap between human instructions and agent actions regarding logging. Finally, the paper offers a lifecycle analysis of post-generation logging regulation, quantifying the hidden maintenance burden placed on human reviewers. We share a replication package~\cite{agentic_logging_rp} which includes our code for conducting the studied experiments, so that others in the research community can replicate or extend our work.


\section{Background \& Related Work}
This paper targets the empirical characterization of logging practices within agentic workflows in Open Source Software (OSS). We analyze how the introduction of AI coding agents impacts the implementation, instruction, and governance of software logging. We discuss the following research directions as they are the closest to our work.

\subsection{Software Logging Practices}
Several studies have characterized how developers implement and maintain logging in real-world systems. Fu et al.~\cite{Fu:2014} analyzed logging in large Microsoft systems and found that logging is highly contextual, typically appearing in specific scenarios such as exception handling, return value verification, and critical logic branches. Pecchia et al.~\cite{Pecchia:2015} examined industrial safety-critical systems, observing that logging practices are largely informal and driven by individual developer expertise rather than standardized guidelines. Yuan et al.~\cite{Yuan:2012} analyzed failure data from distributed systems (e.g., Hadoop) and found critical gaps in logging coverage, noting that many software failures occurred without generating any log entries. Regarding the maintenance of logging code, Kabinna et al.~\cite{Kabinna:2016b} investigated the stability of logging statements in open-source projects, reporting that 20\% to 45\% of logging statements are modified over their lifetime, with many changes occurring shortly after introduction. Li et al.~\cite{LiZ:2021a} highlighted the prevalence of duplication, finding widespread identical static messages that complicate automated analysis. Finally, qualitative studies by Li et al.~\cite{Li:2020} and Rong et al.~\cite{Rong:2023} confirmed that while developers view logging as indispensable for debugging, they struggle with the trade-offs regarding code complexity and performance overhead.

Our research extends this direction by investigating whether coding AI agents adhere to these established human patterns. While prior work characterizes human logging as an informal and unstable activity, it is unknown if AI agents replicate these behaviors (e.g., similar churn rates or coverage gaps) or exhibit distinct ``machine-native'' logging practices. We address this by investigating agentic logging practices against human baselines.

\subsection{LLMs for Software Engineering}
The application of Large Language Models (LLMs) has expanded to cover tasks ranging from code completion to the automation of more complex activities, including aspects of non-functional requirements (NFRs) such as observability. Mastropaolo et al.~\cite{Mastropaolo:2022} introduced LANCE, a T5-based model that treats logging as a translation task. While it achieved 65.9\% accuracy in placement, it struggled with semantic content, achieving only a 15.2\% exact match for log messages. Xu et al.~\cite{UniLog:2024} advanced this work with UniLog, demonstrating that in-context learning and few-shot prompting can significantly improve message quality (BLEU-4 score of 27.1) without the cost of fine-tuning. However, recent empirical evaluations by Rodriguez et al.~\cite{Rodriguez:2025}, utilizing GPT-4o, revealed a persistent bias toward ``over-logging.'' They found that while modern LLMs match human placement accuracy in approximately 64\% of cases, they exhibit an over-logging rate of nearly 83\%, often placing redundant instrumentation at the start or end of functions. Beyond logging, Licorish et al.~\cite{Licorish:2025} observed that while LLMs produce functionally correct code, they frequently introduce verbose structures with higher cyclomatic complexity. Additionally, Sandoval et al.~\cite{Sandoval:2023} identified that LLMs are prone to reproducing insecure patterns present in their training data, such as hard-coded credentials.

Our research complements these benchmark-driven studies with an in-situ analysis of how logging is actually produced by AI agents in real pull requests. While prior evaluations rely on isolated datasets (e.g., LANCE, UniLog), it remains unexamined how these ``over-logging'' and verbose tendencies manifest in active agentic workflows where humans must review and merge the code. We address this by characterizing agent-generated logging in real-world software projects.

\subsection{AI-Assisted Development and Agentic Contributions in OSS}
Recent empirical studies have begun to characterize the growing footprint of AI-generated contributions in open-source ecosystems. Watanabe et al. \cite{Watanabe:2024} analyzed 567 pull requests generated by the Claude Code agent, reporting an acceptance rate of 83.8\%  (comparable to human contributors) while noting that agents primarily focused on maintenance tasks such as refactoring and documentation, with 54.9\% of PRs merged without human modification. He et al.~\cite{He:2025} conducted a study of the Cursor assistant and found that although adoption yielded a transient 3 to 5 times increase in development activity, it coincided with a persistent 30\% rise in static-analysis warnings and a 41\% increase in code complexity, highlighting a trade-off between speed and quality. Wang~et al.~\cite{Wang:2025} identified a ``programmatic bias'' in agentic workflows, observing that agents resort to code-based solutions for 93.8\% of tasks, often diverging from the GUI-driven workflows preferred by human developers. Finally, Tufano et al.~\cite{Tufano:2024} examined developer interactions with LLM-based bots in review processes, finding that while bots are frequently delegated review responsibilities, developers remain skeptical of their suggestions for non-trivial logic changes.

Our work differs from this line of research in that, rather than focusing on functional correctness, code structure, or acceptance rates, we study logging as a mechanism for achieving the non-functional requirement of observability. Specifically, we analyze how logging is produced, explicitly instructed, and regulated within AI-authored pull requests in open-source projects.
\section{Data Collection and Processing}
\label{sec:data_collection}

As the goal of our paper is to understand whether AI agents introduce logging instructions in the same way as human developers, we leverage the AIDev dataset~\cite{Li:2025} that contains human and agentic Pull Requests (PRs). From the dataset, we select a set of repositories along with their agentic and human PRs (Section~\ref{sec:project_selection}). From these PRs, we study the logging statements identified using a keyword-based approach (Section~\ref{sec:logging_identification}). To better understand the agentic behavior in PRs, we analyze how developers instruct agents in terms of the creation and maintenance of logging statements (Section~\ref{sec:llm_jury}).  

\subsection{Repository and PR Selection}
\label{sec:project_selection}

\begin{figure}[t]
\centering
\includegraphics[width=0.7\textwidth,keepaspectratio]{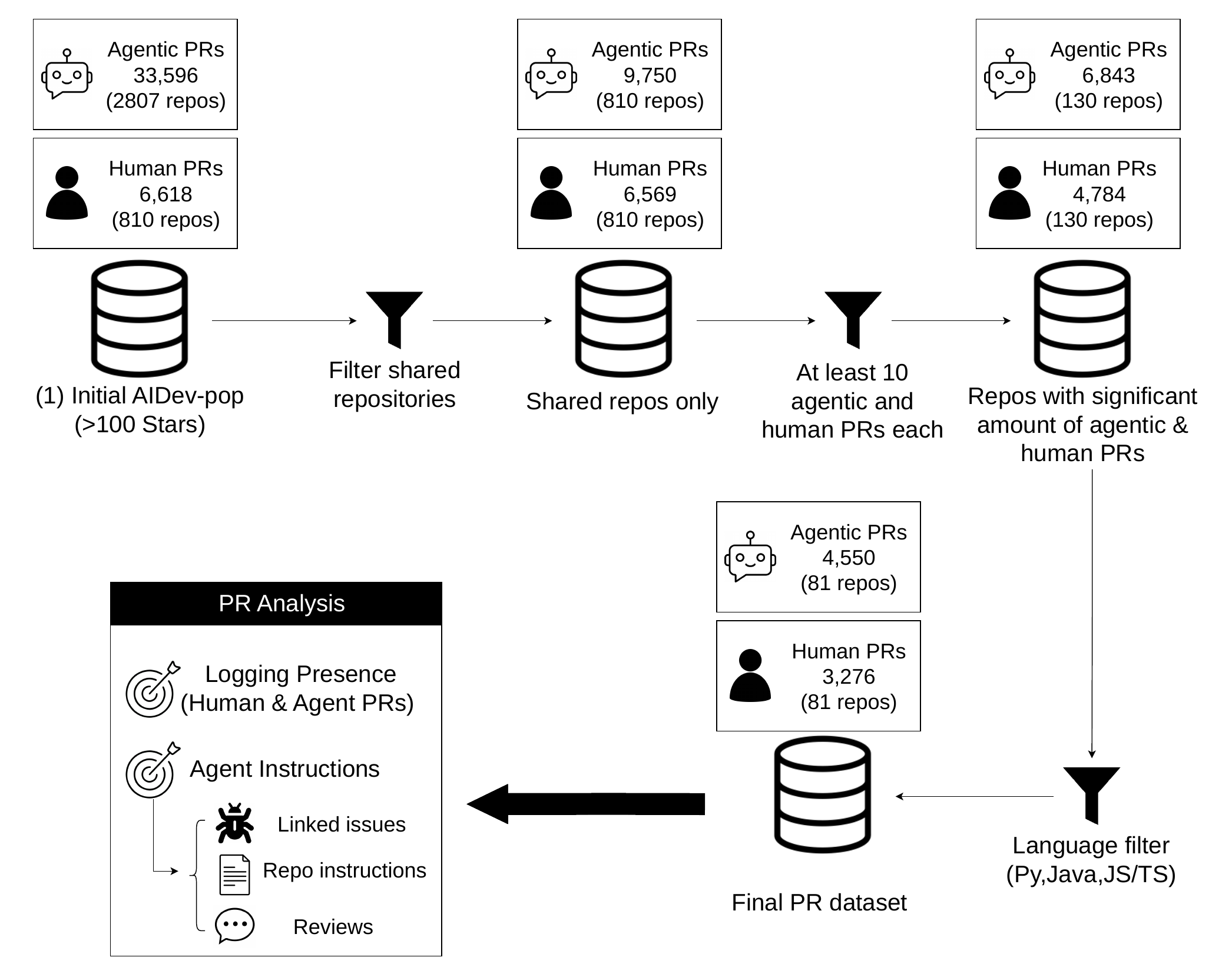}
\caption{Overview of our data collection pipeline.}
\label{fig:datacollection}
\end{figure}

Our data collection pipeline, shown in Figure~\ref{fig:datacollection}, results in a set of 4,550 agentic and 3,276 human PRs across 81 repositories. This dataset is obtained from the AIDev dataset between December 2024 and July 2026. We leverage AIDev-pop, a subset of the AIDev dataset that includes repositories with at least 100 stars, comprising 33,596 agentic PRs and 6,618 sampled human PRs. We further apply a filter restricting to repositories with at least 500 stars to ensure that repositories contain both human and agentic PRs for project-level comparison. The repositories that are in the intersection of the two datasets account for 810 repositories, which together consists of 9,750 agentic and 6,569 human PRs. 

To enable a sound comparison between human and agentic PRs at the project level, we focus on repositories with at least 10 agentic PRs and 10 human PRs. This filtering step yields 130 repositories, with 6,843 agentic and 4,784 human PRs. We further restrict the dataset to repositories whose primary programming language is Python, Java, or JavaScript/TypeScript, similar to prior work on software logging~\cite{Chou:2025,Li:2017,LiZ:2021a,Ouatiti:2024,Ouatiti:2023}. We focus on these programming languages since they have well-defined strategies for identifying the logging statements (as further discussed in Section~\ref{sec:logging_identification} below).


For each PR, we collect its patch to determine whether it includes changes to logging statements. For each agentic PR, we also collect the associated instructions. To identify PRs with logging changes, we use the GitHub API to retrieve patches for human PRs, while patches for agentic PRs are already available in the AIDev dataset. Further details on logging statement identification are provided in the next subsection. As our study covers how agents are instructed, we further collect the instructions given to the agents. These instructions can be in the form of issues linked to PRs, repository-level instructions files (e.g., \textit{./github/copilot-instructions.md}) at the time of the PR creation, or comments provided during the review for agents to adjust their generated code. Note that instructions can also be provided through other channels that are not publicly available for collection and analysis.

\subsection{Logging Detection Strategy}
\label{sec:logging_identification}
We identify logging statement changes within code diffs using a regex-based strategy adapted from prior logging studies~\cite{Li:2017,LiZ:2021a,Ouatiti:2024, Ouatiti:2023}. As detailed in Tables~\ref{tab:regex_logging_patterns} and~\ref{tab:regex_logging_scope}, we use regex expressions tailored to each programming language. These expressions are executed on source files based on their extensions, such as \textit{.py} for Python. We explicitly exclude build artifacts (e.g., \texttt{dist/}, \texttt{node\_modules/}), binary assets, and minified code to reduce noise from auto-generated files. The regex patterns capture logging framework invocations in Python (e.g., \texttt{logging.info}), object-oriented styles in Java and JavaScript (e.g., \texttt{LOGGER.warn}), and console logging in JavaScript/TypeScript (e.g., \texttt{console.log}). Generic print statements (e.g., \texttt{System.out.println}) are excluded, as they do not represent typical production-level logging.

To ensure the robustness of our regex expressions, we perform a manual analysis of a representative sample of 380 diffs (95\% confidence interval and 5\% margin of error) and find that they achieve a precision of 96\% and a recall of 94\%, demonstrating the reliability of our regex-based approach. 

Note that we exclude 4 repositories in which neither agentic nor human PRs contain logging changes, resulting in a final dataset of 77 repositories.


    
    

\begin{table}[t]
\centering
\caption{Regex expressions used to identify logging statements (case-insensitive; \texttt{re.I}).}
\label{tab:regex_logging_patterns}
\small
\setlength{\tabcolsep}{4pt}
\begin{tabularx}{0.7\linewidth}{l X}
\toprule
\textbf{Language} & \textbf{Regex Pattern} \\
\midrule
Python &
\texttt{\textbackslash b(?:logging|logger|\_logger)\textbackslash .} \newline
\texttt{(?P<level>debug|info|warning|warn|error|} \newline
\texttt{critical|exception)\textbackslash s*\textbackslash (} \\
\addlinespace
Java &
\texttt{\textbackslash b(?:logger|log|LOG|LOGGER)\textbackslash .} \newline
\texttt{(?P<level>trace|debug|info|warn|warning|} \newline
\texttt{error|fatal)\textbackslash s*\textbackslash (} \\
\addlinespace
JS/TS &
\texttt{\textbackslash bconsole\textbackslash .} \newline
\texttt{(?P<level>log|info|warn|error|debug)\textbackslash s*\textbackslash (} \\
\bottomrule
\end{tabularx}
\end{table}

\begin{table}[t]
\centering
\caption{File extensions scanned and path/suffix exclusions used to reduce noise during logging statement identification.}
\label{tab:regex_logging_scope}
\small
\setlength{\tabcolsep}{4pt}
\begin{tabularx}{\linewidth}{l p{3.2cm} X}
\toprule
\textbf{Language} & \textbf{Extensions} & \textbf{Exclusions (Paths \& Suffixes)} \\
\midrule
Python & \texttt{.py} &
\textbf{Dirs:} \texttt{build/}, \texttt{dist/}, \texttt{site-packages/}, \texttt{vendor/} \newline
\textbf{Files:} \texttt{\_test.py}, \texttt{test\_*.py} \\
\addlinespace
Java & \texttt{.java} &
\textbf{Dirs:} \texttt{target/}, \texttt{bin/}, \texttt{build/} \newline
\textbf{Files:} \texttt{Test.java}, \texttt{*Test.java} \\
\addlinespace
JS/TS & \texttt{.js}, \texttt{.jsx}, \newline \texttt{.ts}, \texttt{.tsx} &
\textbf{Dirs:} \texttt{node\_modules/}, \texttt{dist/}, \texttt{public/}, \texttt{vendor/} \newline
\textbf{Suffixes:} \texttt{.min.js}, \texttt{.map}, \texttt{.gz}, \texttt{.bundle.js}, \texttt{.worker.js} \\
\bottomrule
\end{tabularx}
\end{table}





\subsection{Studying Agent Instructions}
\label{sec:llm_jury}

We collect the available instructions for the studied repositories at the creation time of each individual agentic PR, as discussed in Section~\ref{sec:agentic_instructions_collection}. We then leverage an LLM-as-judge multi-agent approach to identify the logging intent of developers (e.g., creation of a new logging statement), as discussed in Section~\ref{sec:intent_identification}.

\subsubsection{Collection of Agent Instructions}
\label{sec:agentic_instructions_collection}

In this paper, we study how developers instruct agents to generate and maintain logs. The instruction dataset consists of three sources, as described below.
\begin{itemize} 
\item Linked issues: Developers can create an issue describing a task and assign it to an agent, which then addresses the task through a PR. Issues associated with agentic PRs are available in the AIDev dataset and are used to analyze how developers provide logging-related instructions. Not all agentic PRs have associated issues, as agents may co-author PRs offline with human developers who subsequently submit them. 

\item Repository-level agents' instruction files: Developers can guide agents through repository-level instruction files (e.g., \textit{CLAUDE.md} and \textit{.github/copilot-instructions.md}). For each agentic PR, we retrieve the instruction files present in the repository at the time the PR was created. These files are identified using the regex patterns shown in Table~\ref{tab:instruction_regex}. 

\item Review comments on agentic PRs: Developers can leave review comments under a PR to guide agents. We collect review comments available in the AIDev dataset as a means of capturing instructions provided to the AI agent to adjust its generated code.
\end{itemize}

\begin{table}[t]
\centering
\caption{Agent instruction files and their corresponding regex patterns used for identification.}
\label{tab:instruction_regex}
\small
\begin{tabular}{l p{10.5cm}}
\toprule
\textbf{Agent} & \textbf{Regex pattern} \\
\midrule
Devin  & \texttt{**/PULL\_REQUEST\_TEMPLATE/DEVIN\_PR\_TEMPLATE.md}; \texttt{**/PULL\_REQUEST\_TEMPLATE/devin\_pr\_template.md} \\
Cursor & \texttt{.cursor/*}; \texttt{.cursorrules}; \texttt{**/*.mdc} \\
Copilot & \texttt{.github/copilot-instructions.md}; \texttt{.github/instructions/*} \\
Claude & \texttt{.claude/*}; \texttt{CLAUDE.md}; \texttt{.github/workflows/claude*.yml} \\
Codex  & \texttt{**/AGENTS.override.md}; \texttt{**/TEAM\_GUIDE.md}; \texttt{**/.agents.md} \\
Common & \texttt{**/AGENTS.md} \\
\bottomrule
\end{tabular}
\end{table}



\subsubsection{Identifying Developers' Logging-Related Intents}
\label{sec:intent_identification}

To identify the intents of developers behind logging instructions (i.e., Add, Remove, or Modify), if any, we use a multi-agent LLM-as-judge protocol~\cite{Li:2024,Hao:2025,Verga:2024} on review comments, instruction files, and associated issues. Each of these data points is studied separately to identify whether it has a logging instruction. If so, which of the three possible typical intents (Add, Remove, or Modify) is provided. To do so, we prompt three frontier models (GPT-4o, GLM-4.7, and DeepSeek-V3.2) to independently classify a text input (e.g., a code review comment) using the prompt shown in Figure~\ref{fig:jury_prompt_issues}, whose construction is discussed below. From the three votes, we assign the final label for each instruction source (e.g., a review comment) by majority voting.

To construct our prompt, we follow a similar approach to previous work~\cite{Hao:2025,Watanabe:2025}. We first establish ground truth labels for 100 samples through manual annotation. Using this sample, we iteratively refine the jury prompt, measuring the agreement between the jury's majority-vote prediction and our manually curated ground truth. We finalize the prompt once this agreement reaches Cohen's $\kappa = 0.83$, ensuring that the automated classification reliably mirrors human intent.


\begin{figure}[t]
\centering
\begin{tcolorbox}[colback=gray!5, colframe=black, title=\textbf{LLM Jury Prompt for Instructions (Issues, Reviews \& Repo Files)}]
\textbf{System Message:} \\
You are an expert reviewer analyzing text files for logging-related directives. The text file could either be an issue body, a repository-level agent instruction file, or a PR review comment.
\vspace{2mm}
\hrule
\vspace{2mm}
\textbf{User Message Template:} \\
Task: Analyze the provided text. Does it contain an explicit instruction regarding logging, tracing, observability?

Output Labels (Choose one):
\begin{itemize}
    \item \texttt{ADD}: Requests for new logs (quantity control). Examples: ``Add debug logs'', ``Log all errors'', ``Ensure observability'', ``Include logging statements'', ``Always log exceptions''.
    \item \texttt{REMOVE}: Requests to suppress logs (noise control). Examples: ``No console.log in production'', ``Reduce verbosity'', ``Don't add debug logs'', ``Avoid excessive logging''.
    \item \texttt{MODIFY}: Requests to modify existing logs (quality control). Examples: ``Use slf4j instead of System.out'', ``Change info to debug level'', ``Use structured logging format'', ``Follow our logging conventions''.
    \item \texttt{none}: No specific logging instructions found.
\end{itemize}
Important:
\begin{itemize}
    \item Only classify as ADD, REMOVE, or MODIFY if there is an EXPLICIT instruction about logging behavior.
    \item Vague mentions of ``good practices'' or ``code quality'' without specific logging guidance should be NONE.
    \item If the instruction says to USE logging (e.g., ``Use console.log for debugging''), that is ADD.
    \item If the instruction specifies HOW to log but not whether to add/remove, that is MODIFY.
\end{itemize}
Return a compact JSON object: \texttt{\{``label'': ``...'', ``rationale'': ``...''\}}.

Type of text: \texttt{\{metadata\}}

Text: \\
\texttt{\{text\}}
\end{tcolorbox}
\caption{The prompt used to identify logging instructions. The \texttt{metadata} injection provides the model with the source type (i.e., \texttt{issue}, \texttt{repo\_instruction}, or \texttt{review\_comment})}
\label{fig:jury_prompt_issues}
\end{figure}
\section{Results}
\label{sec:results}
\subsection*{\textbf{RQ1. \RQI}}

\textbf{Motivation:} The goal of this research question is to determine whether AI agents mimic human developers in the creation and maintenance of logs within a given project. In other words, if humans frequently insert logs in a particular way, do agents follow the same logging practices? Understanding this behavior helps identify whether agents are capable of automatically recognizing and following developers’ practices in creating logging statements, or whether they neglect such practices, potentially diminishing the observability of a software system. The results of this research question motivate the need to equip AI agents with tools that analyze and respect project-specific observability practices, if agents do not follow human logging conventions, and to encourage developers to be more explicit about their logging expectations.

\begin{table}[t]
\centering
\caption{Log levels categorized by verbosity across analyzed languages.}
\label{tab:log_levels}
\small
\begin{tabular}{clll}
\toprule
\textbf{Verbosity} & \textbf{Python} & \textbf{Java} & \textbf{JS/TS} \\
\midrule
\textit{Highest} & \texttt{debug} & \texttt{trace}, \texttt{debug} & \texttt{debug} \\
$\downarrow$ & \texttt{info} & \texttt{info} & \texttt{info}, \texttt{log} \\
$\downarrow$ & \texttt{warning} & \texttt{warn} & \texttt{warn} \\
\textit{Lowest} & \texttt{error}, \texttt{critical} & \texttt{error}, \texttt{fatal} & \texttt{error} \\
\bottomrule
\end{tabular}
\end{table}

\begin{table}[t]
\centering
\caption{Mapping of language-specific keywords to unified syntactic contexts.}
\label{tab:syntactic_mapping}
\small
\begin{tabular}{lp{3.5cm}p{3.5cm}}
\toprule
\textbf{Unified Context} & \textbf{Python Keywords} & \textbf{Java \& JS/TS Keywords} \\
\midrule
\textbf{Conditionals} & \texttt{if}, \texttt{elif}, \texttt{else} & \texttt{if}, \texttt{else}, \texttt{switch}, \texttt{case} \\
\textbf{Loops} & \texttt{for}, \texttt{while} & \texttt{for}, \texttt{while}, \texttt{do} \\
\textbf{Try/Catch} & \texttt{try}, \texttt{except}, \texttt{finally} & \texttt{try}, \texttt{catch}, \texttt{finally} \\
\textbf{Unnested} & \multicolumn{2}{l}{\textit{Logging statement is not enclosed in any of the above blocks}} \\
\bottomrule
\end{tabular}
\end{table}

\smallskip\noindent
\textbf{Approach:} To identify whether AI agents mimic developers in the creation and maintenance of logging statements, we compare agentic PRs with human PRs within the same project. For each of the 77 repositories in our dataset, we calculate the following metrics to characterize logging practices for both agentic and human PRs:

\begin{itemize} 
    \item Logging Prevalence: We measure the percentage of PRs that explicitly introduce, modify, or remove at least one logging statement. This metric is computed separately for human and agentic PRs. 
    \item Log Density: The number of modified logging statements per 1,000 modified lines of code (LOC). 
    \item Message Characteristics: We study message verbosity and log levels. Verbosity is measured as the number of characters in extracted literal log message text. Log level distributions are computed using language-specific logging patterns (Table~\ref{tab:log_levels}). At the repository level, log message length comparisons are performed only when both agentic and human PRs contain at least one extractable log message. Consequently, repositories without extractable messages on at least one side (e.g., those with only variable-based or dynamically constructed messages) are excluded from the verbosity analysis. For project-level comparisons, we use the median log message length and the median log level percentages computed across PRs.
    \item Syntactic Context: The distribution of log placement within control-flow constructs (e.g., \texttt{if}, \texttt{try/catch}, and \texttt{Unnested}). To account for language differences in our diff-based analysis, we map language-specific keywords found in the diff context to unified categories, as shown in Table~\ref{tab:syntactic_mapping}. For example, both Python's \texttt{except} and Java's \texttt{catch} are mapped to the unified \texttt{try/catch} category. For project-level metrics, we calculate the median across all PRs of that project. For each control-flow category (e.g., try/catch), we calculate the median percentage of logs placed in that construct across all PRs in the project. 
\end{itemize}

We focus on these metrics to characterize logging practices, as they capture the main aspects of logging statements and align with prior literature on the development and maintenance of software logging~\cite{Mohammed:2024}. Maintenance effort, log density, verbosity, log level distribution, and syntactic context have all been studied in prior work~\cite{Mohammed:2024,Foalem:2024,Li:2017,LiZ:2021a}. 

For each project and PR type (i.e., human or agentic), we calculate one project-level metric value as the median across all PRs of that type (e.g., median logging prevalence across all human PRs, and separately across all agentic PRs). We compare agentic to human PRs using a normalized score computed as $\frac{\textit{Agentic}}{\textit{Agentic} + \textit{Human}}$. This score maps each repository to a common 0--1 scale and remains defined as long as at least one side is non-zero. A score of 0.5 indicates parity, values above 0.5 indicate higher agentic values, and values below 0.5 indicate higher human values.

\begin{figure}[t]
\centering
\begin{subfigure}[t]{0.48\textwidth}
    \centering
    \includegraphics[width=\textwidth,keepaspectratio]{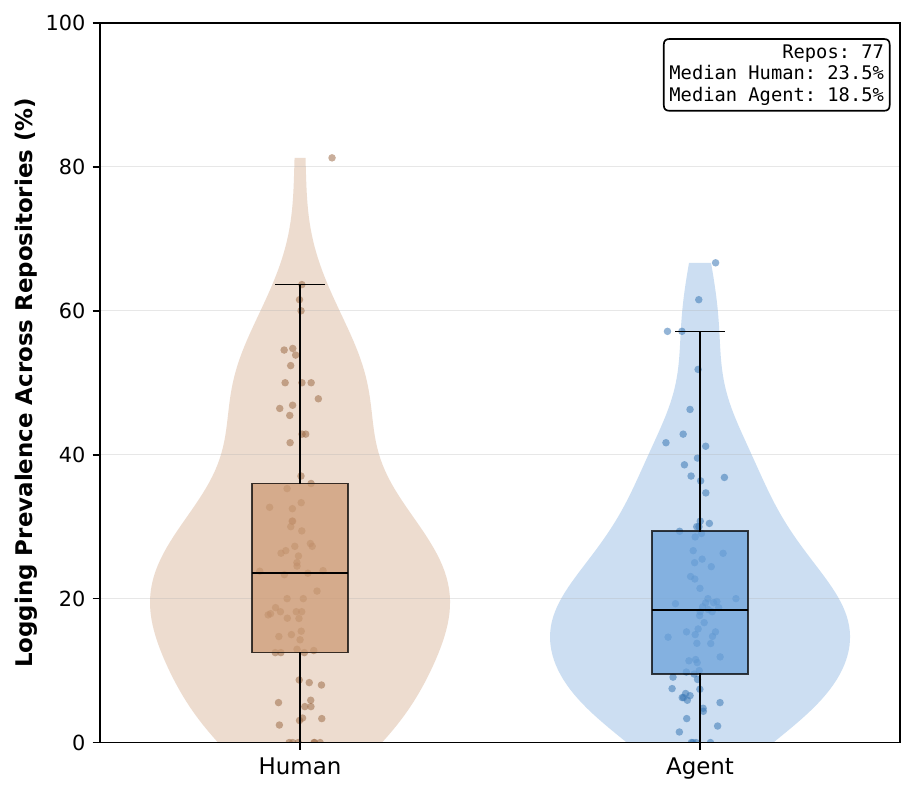}
    \caption{Distribution of repository-level logging prevalence for human and agentic PRs.}
    \label{fig:prevalence_dist}
\end{subfigure}
\hfill
\begin{subfigure}[t]{0.48\textwidth}
    \centering
    \includegraphics[width=\textwidth,keepaspectratio]{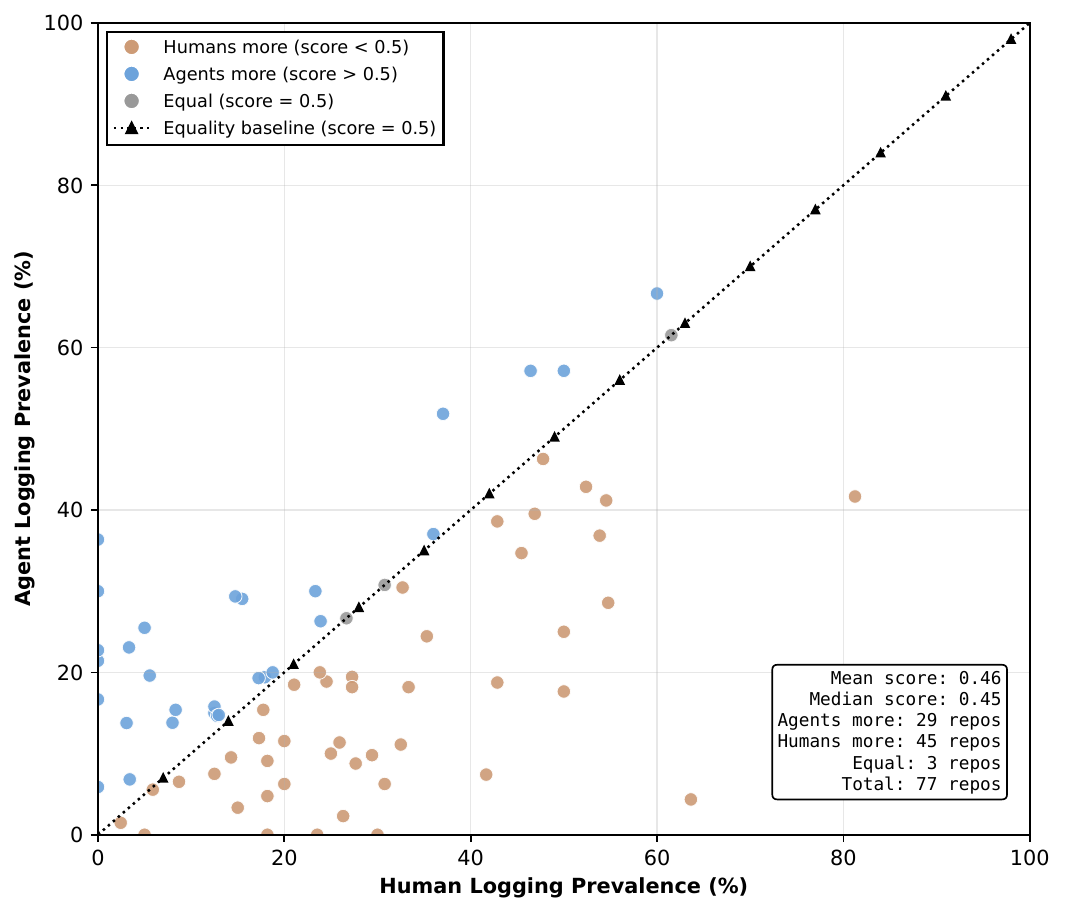}
    \caption{Paired repository-level comparison of logging prevalence (Human on x-axis, Agent on y-axis).}
    \label{fig:prevalence_scatter}
\end{subfigure}
\caption{Repository-level comparison of logging prevalence in human and agentic PRs. Panel (a) shows the distribution across repositories; panel (b) shows the paired per-repository comparison.}
\label{fig:prevalence_ratio}
\end{figure}

\smallskip\noindent
\textbf{Results:} \textbf{In 58.4\% of the studied repositories, human pull requests change logging (i.e., add, modify, or remove) more often than agent pull requests}, as shown in Figure~\ref{fig:prevalence_ratio}. This means that, within the same repository, humans are more likely to add, remove, or modify logging statements when they change code. Specifically, in 45 out of 77 repositories~(58.4\%), the agentic-to-human logging prevalence score is below 0.5, indicating that agents touch logging in a smaller share of their PRs than humans do. In contrast, 29 repositories (37.7\%) show the opposite trend, where agents touch logging more often than humans. This difference between agent and human logging prevalence across the same repositories is statistically significant ($p=0.019$). Moreover, the median score is 0.45, suggesting that for a typical project in our dataset, agents change logging about 16\% less often than humans. Finally, we observe that among the 22 repositories~(28.6\%) with similar logging prevalence scores (from 0.44 to 0.55), logging prevalence varies substantially, ranging from 5.6\% to 66.7\%, with medians of 26.5\% for agents and 25.3\% for humans.

\begin{figure}[t]
\centering
\begin{subfigure}[t]{0.48\textwidth}
    \centering
    \includegraphics[width=\textwidth,keepaspectratio]{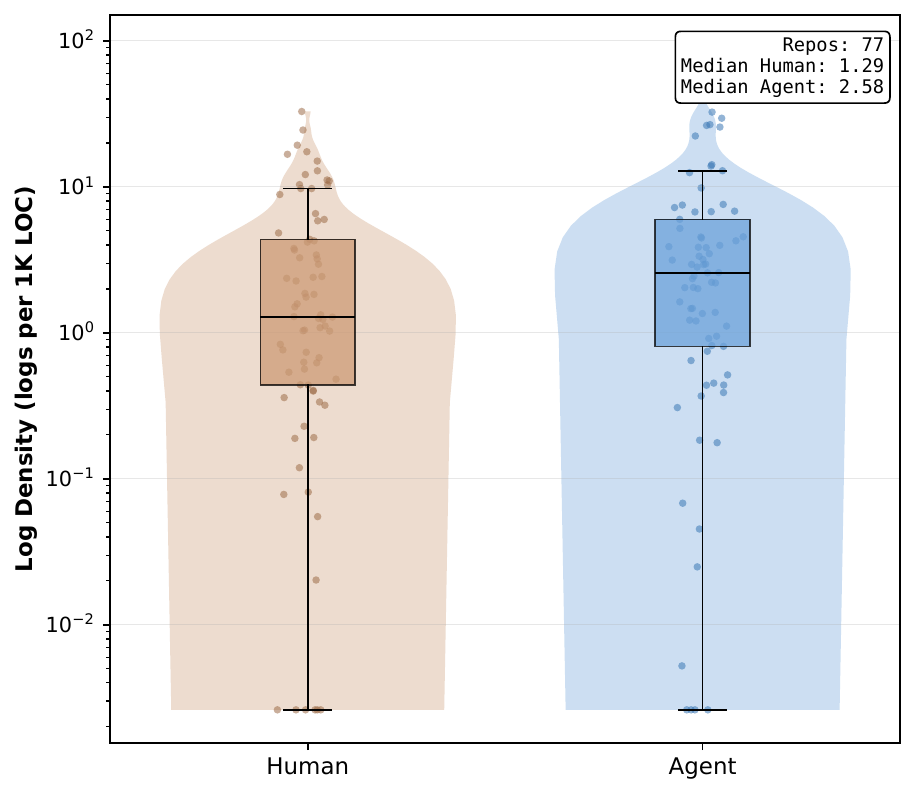}
    \caption{Distribution of repository-level log density for human and agentic PRs.}
    \label{fig:log_density_dist}
\end{subfigure}
\hfill
\begin{subfigure}[t]{0.48\textwidth}
    \centering
    \includegraphics[width=\textwidth,keepaspectratio]{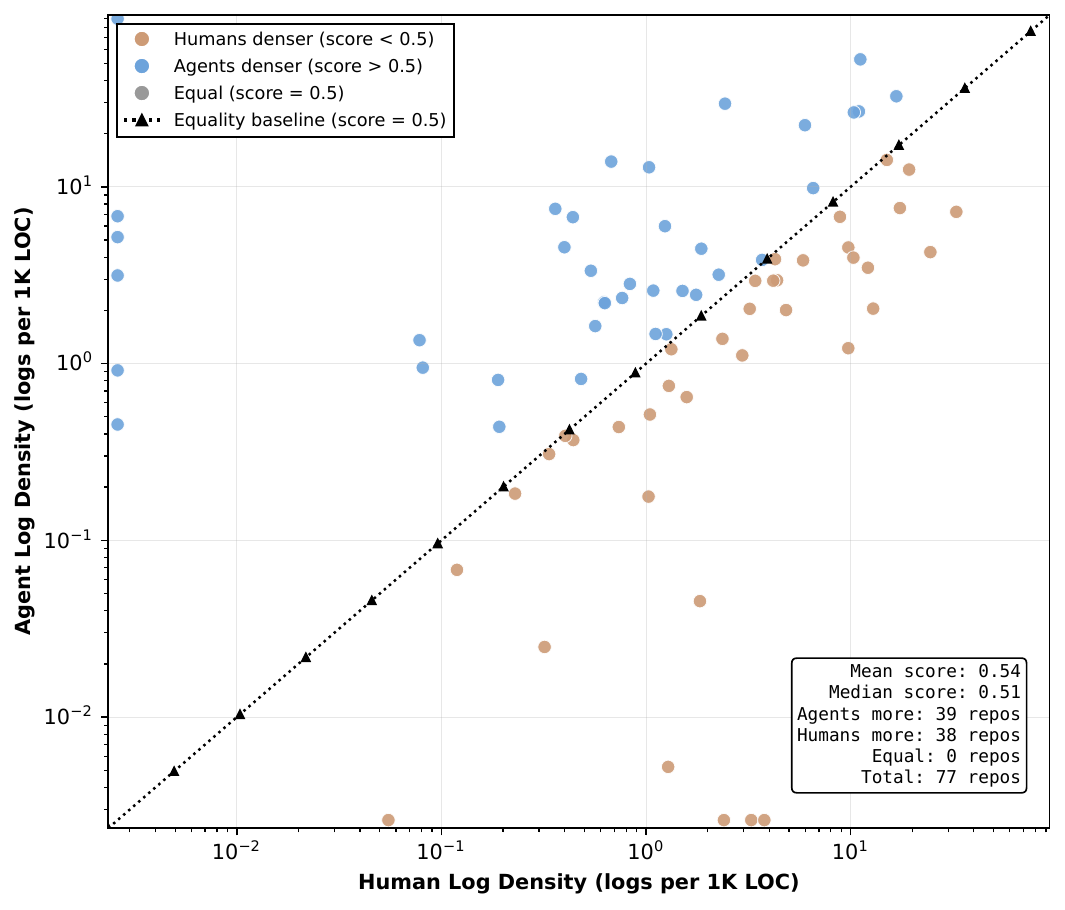}
    \caption{Paired repository-level comparison of log density (Human on x-axis, Agent on y-axis).}
    \label{fig:log_density_scatter}
\end{subfigure}
\caption{Repository-level comparison of log density in human and agentic PRs. Panel (a) shows the distribution across repositories; panel (b) shows the paired per-repository comparison.}
\label{fig:log_density_ratio}
\end{figure}

\textbf{In 50.6\% of the studied repositories, agentic pull requests have higher log density than human pull requests}, as shown in Figure~\ref{fig:log_density_ratio}. However, the paired repository-level difference in log density is not statistically significant ($p=0.274$). The median agentic-to-human density score is 0.51. Consequently, in a typical project, agent and human log density are nearly balanced, with a slight tilt toward agents. For example, in \textit{microsoft/ApplicationInsights-JS} the mean log density is 12.90 for agents versus 1.03 for humans (score 0.93). Restricting the comparison to the 67 repositories~(87.0\%) where both agents and humans have logging-changing PRs reveals a stronger pattern as the median score rises to 0.56, which means that, in repositories where both sides actively add logs, agents produce about 30\% more logging changes per 1,000 changed LOC than humans.

\begin{figure}[t]
  \centering
  \includegraphics[width=0.5\textwidth,keepaspectratio]{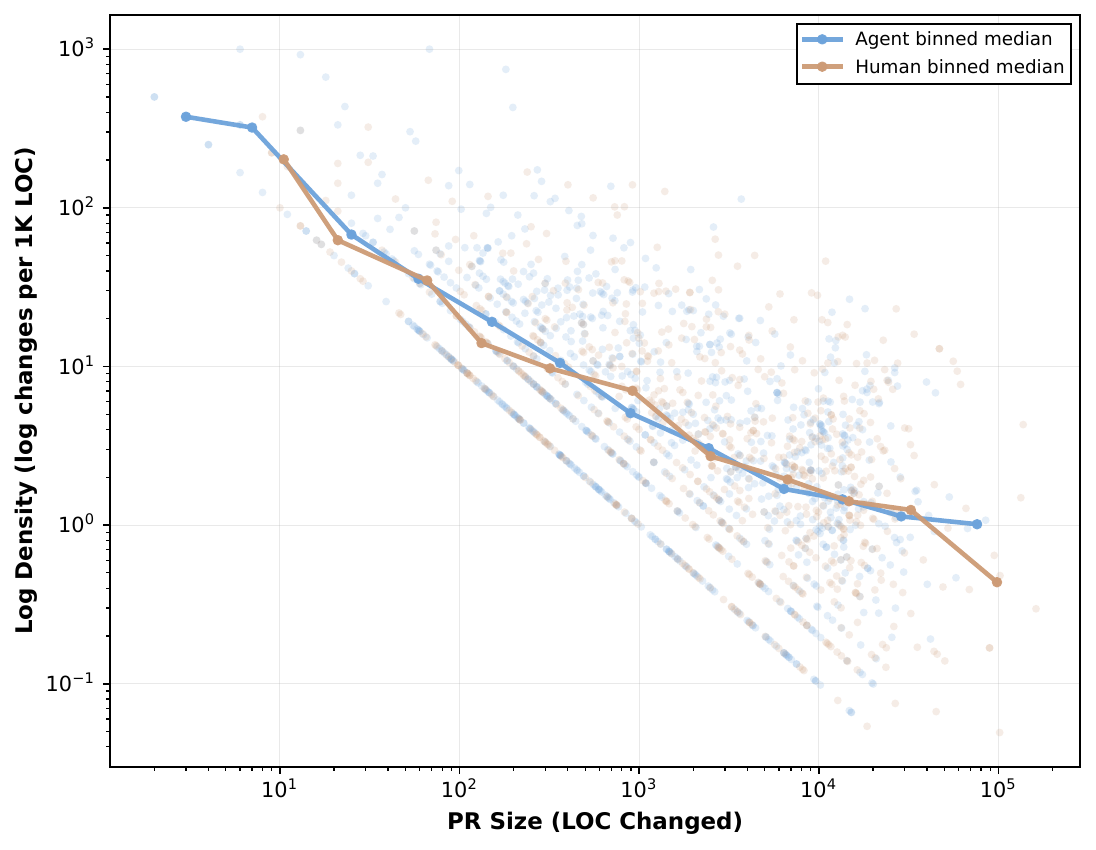}
  \caption{PR-level relationship between change size and log density. Points represent individual PRs (transparent). Lines represent binned medians: PR size is partitioned into 12 log-spaced bins over the combined agent and human LOC range, and for each group we plot the median LOC (x-axis) and median log density (y-axis).}
  \label{fig:density_vs_loc}
  \end{figure}

However, as illustrated in Figure~\ref{fig:density_vs_loc}, this density gap is largely a composition effect across PR sizes rather than a fundamental difference in logging behavior. Because log density naturally decreases as PR size increases for both groups, the overall density metrics are skewed by the fact that agents typically make much smaller modifications (median 1,279 LOC versus 2,770.5 LOC for humans). Agent log-adding PRs are heavily concentrated in smaller ranges (46.8\% for agents vs.\ 33.0\% for humans for LOC $\leq$1,000), which naturally yield denser patches. Indeed, in the 48 repositories where agents make smaller changes than humans, they add 65\% more logs per 1,000 LOC.

Conversely, human log-adding PRs are concentrated in massive changes (52.7\% for humans vs.\ 40.7\% for agents for LOC $>$2,500). When controlling for this size disparity, the logging behaviors converge. In the 19 repositories where agents make larger changes than humans, agents become more conservative, adding 21\% fewer logs. Similarly, when examining only large PRs ($>$2,500 LOC), median densities between the two groups become nearly identical (1.64 vs.\ 1.59 logs per 1,000 LOC). This pattern strongly supports a dynamic of selective delegation~\cite{Watanabe:2025} as developers predominantly trust agents with smaller, tightly bounded tasks, while humans handle larger architectural integrations.

\begin{figure}[t]
\centering
\begin{subfigure}[t]{0.48\textwidth}
    \centering
    \includegraphics[width=\textwidth,keepaspectratio]{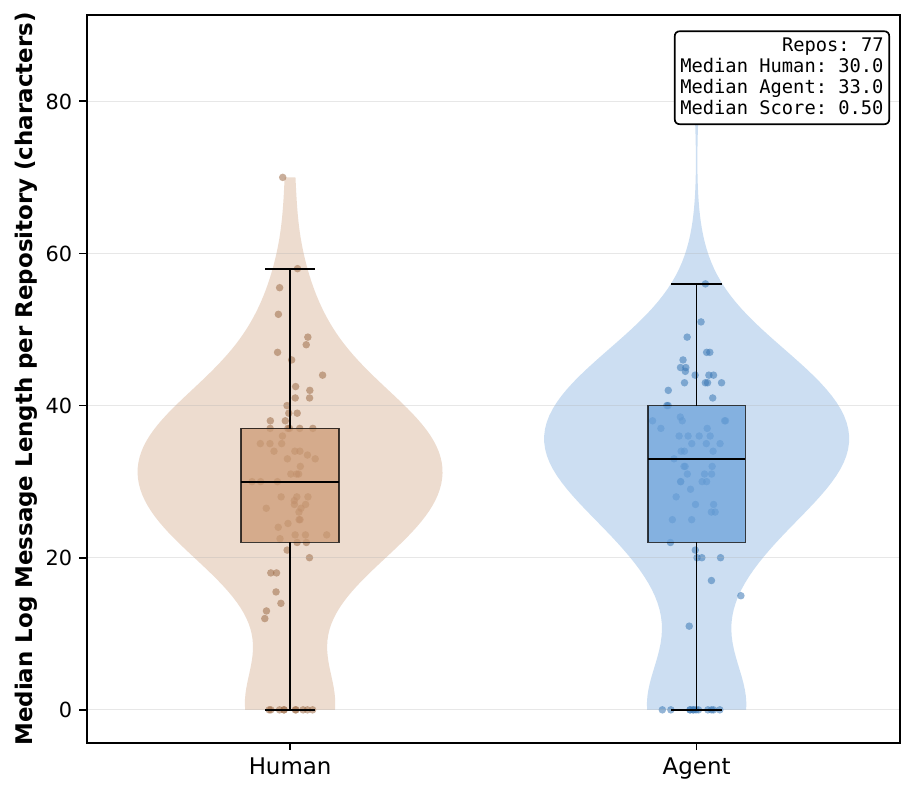}
    \caption{Distribution of repository-level log-message length for human and agentic PRs.}
    \label{fig:verbosity_dist}
\end{subfigure}
\hfill
\begin{subfigure}[t]{0.48\textwidth}
    \centering
    \includegraphics[width=\textwidth,keepaspectratio]{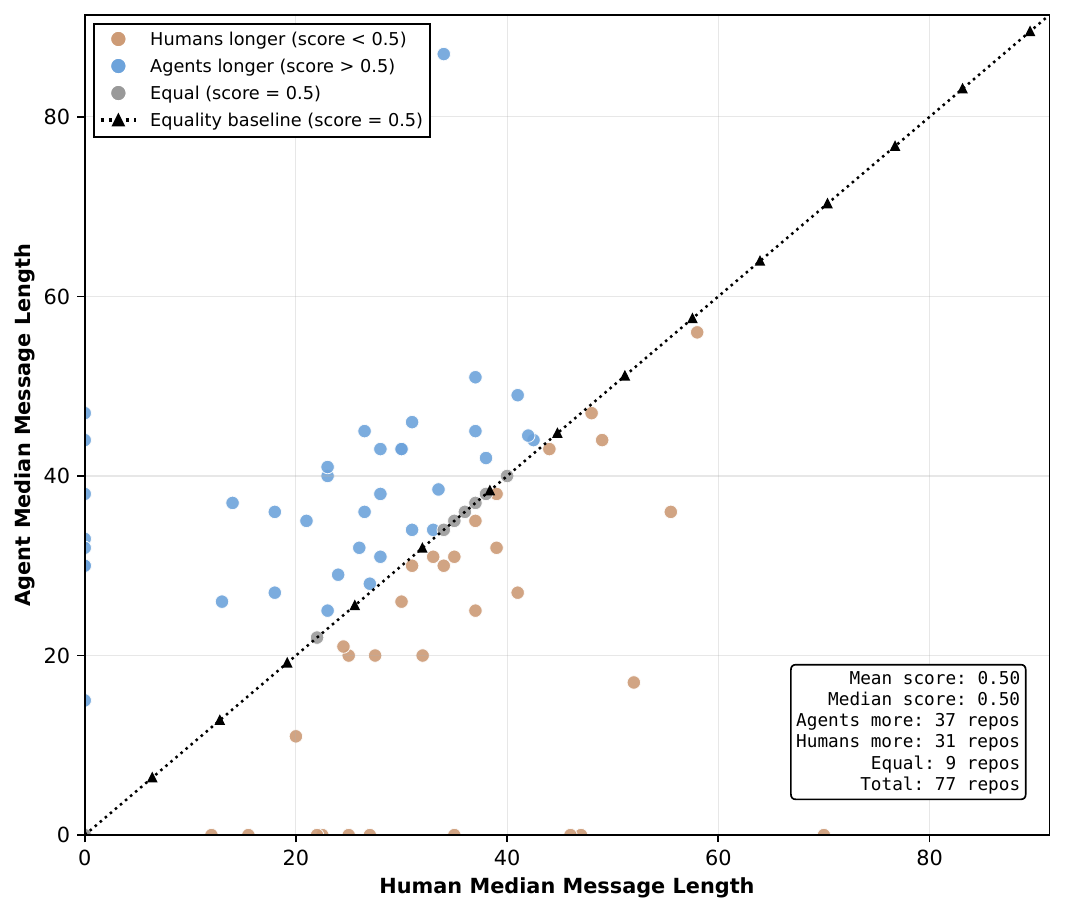}
    \caption{Paired repository-level comparison of log-message length (Human on x-axis, Agent on y-axis).}
    \label{fig:verbosity_scatter}
\end{subfigure}
\caption{Repository-level comparison of log-message length in human and agentic PRs. Panel (a) shows the distribution across repositories; panel (b) shows paired per-repository medians. Results are shown for the same 57 repositories with extractable message text on both sides.}
\label{fig:verbosity_combined}
\end{figure}

\textbf{Agents and humans write log messages of similar length at the repository level}, as shown in Figure~\ref{fig:verbosity_combined}. Across the studied 77 repositories, log message length (measured in number of characters) is centered at parity, with a median score of 0.50. Furthermore, 63.6\% (49 out of 77) of the repositories show similar log message lengths between agents and humans. In contrast, 22.1\% (17 out of 77) of the repositories show agents writing substantially longer messages, while 14.3\% (11 repositories) show humans writing substantially longer messages. For example, \textit{wix/react-native-ui-lib} has a median agentic message length of 37 compared to 14 for humans (score 0.73), while \textit{jina-ai/node-DeepResearch} shows the opposite pattern (17 vs.\ 52, score 0.25). The median score of 0.50 suggests that agents write messages nearly identical in length to humans, indicating they largely follow existing human practices.

\begin{figure}[t]
\centering
\includegraphics[width=0.8\textwidth,keepaspectratio]{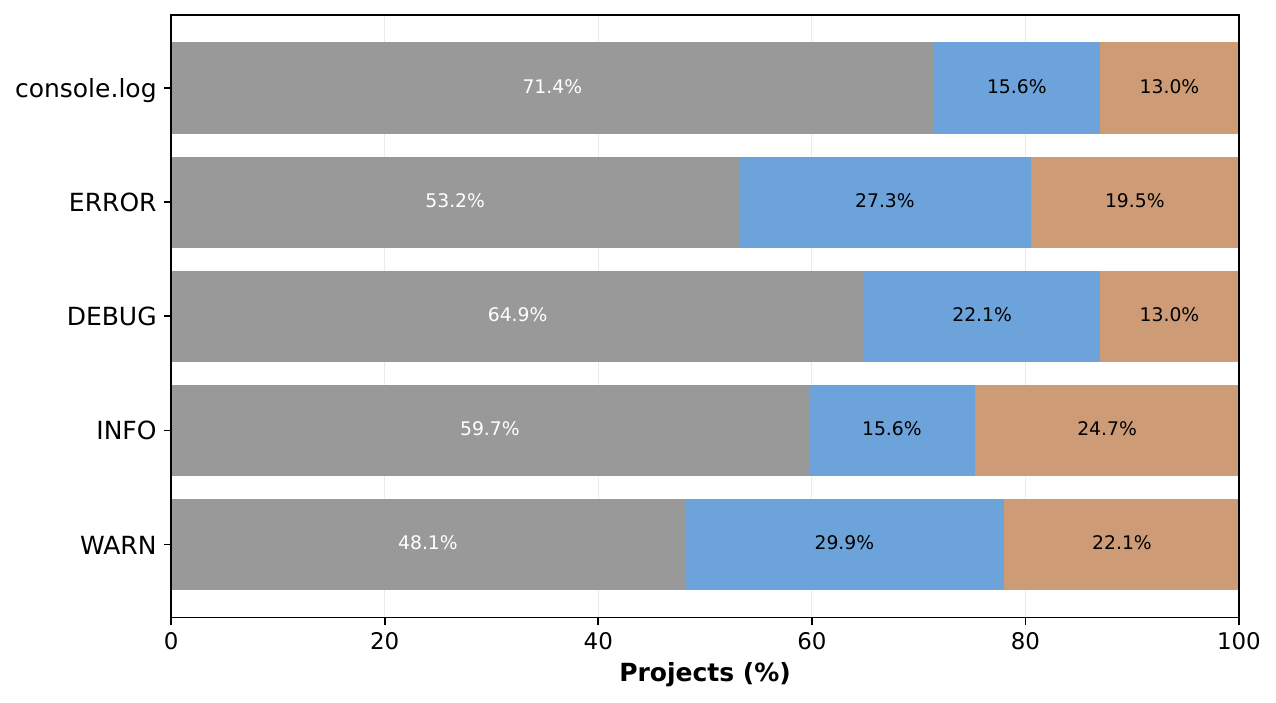}
\caption{Percentage of repositories in three categories for each log level: gray = similar usage, blue = agents use more, and tan = humans use more.}
\label{fig:log_level_dist}
\end{figure}

\textbf{Agents largely mirror human conventions for most log levels, with notable exceptions for \texttt{INFO} and \texttt{WARN} messages}, as shown in Figure~\ref{fig:log_level_dist}. Specifically, agents show high adherence to project norms for general-purpose and error logging. In fact, the usage rates of the standard JS/TS logging method \texttt{console.log}\footnote{Serves as the default, severity-neutral logging method in JS/TS} and the \texttt{ERROR} log level are similar in 71.4\% and 53.2\% of our studied repositories respectively. The \texttt{DEBUG} log level also shows high alignment (similar in 64.9\% of the repositories), with agents using it more in 22.1\% of repositories and humans in only 13.0\%. However, divergence appears with \texttt{INFO} and \texttt{WARN}, as \texttt{INFO} is the level where humans most often exceed agents (24.7\% of repositories), while \texttt{WARN} shows the lowest overall similarity (48.1\%), with agents overusing it in 29.9\% of repositories and humans in 22.1\%. A manual inspection suggests that part of this gap may come from program-state confirmation messages (e.g., \texttt{``operation completed''}), which are more common in human-authored logs.

\begin{figure}[t]
\centering
\includegraphics[width=0.75\linewidth]{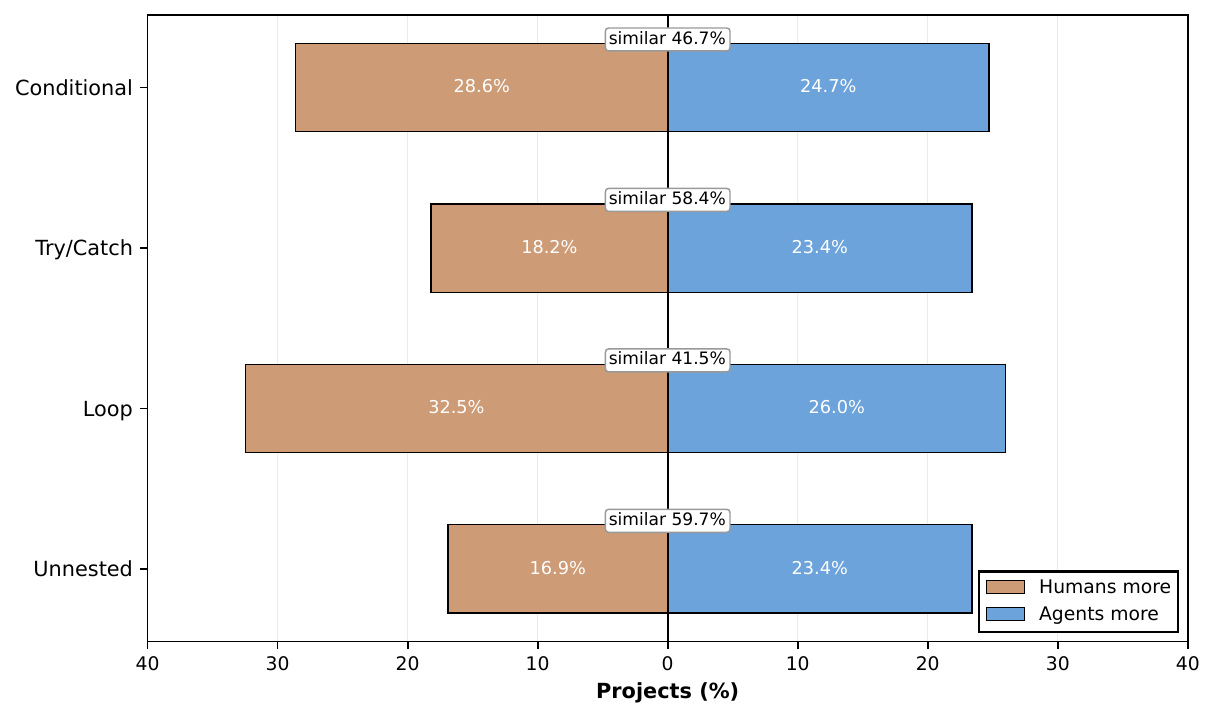}
\caption{Aggregated syntactic-context comparison across repositories. Left bars show repositories where humans use the context more; right bars show repositories where agents use it more; center labels show the share of repositories with similar usage.}
\label{fig:rq1_context_symmetry}
\end{figure}

\textbf{Agents largely mirror human log placement conventions, with divergence in conditional and iterative contexts}, as shown in Figure~\ref{fig:rq1_context_symmetry}. Specifically, agents show high adherence to project norms for error-handling and top-level logging. In fact, the placement of logs in \texttt{TRY\_CATCH} blocks and \texttt{UNNESTED} (top-level function body) contexts are nearly identical in 58.4\% and 59.7\% of our studied repositories respectively. However, divergence appears in control-flow contexts. For instance, we observe for the \texttt{CONDITIONAL} blocks (if/else/switch) that only 46.7\% of repositories show similar usage, with humans placing more logs in conditionals in 28.6\% of repositories. The gap widens for \texttt{LOOP} contexts, where humans log significantly more in 32.5\% of repositories. These log placement patterns resemble the log level findings, as agents match human practices for error-related contexts but are more conservative in locations where informational logging typically occurs (e.g., Loops).

\begin{tcolorbox}[colback=gray!10, colframe=black, title=\textbf{Summary of RQ1}]
In 58.4\% of repositories, agents change logging in fewer PRs than humans (median prevalence score 0.45). Conversely, when both agents and humans do add logs, 58.2\% of repositories show higher agent log density, often associated with smaller change sizes. Moreover, message length, log levels, and syntactic placement are broadly similar between agentic and human PRs. We recommend developers to review logging explicitly in agentic PRs, checking in particular for missing instrumentation and overly dense logging in small PRs.
\end{tcolorbox}

\subsection*{\textbf{RQ2. \RQII}}

\textbf{Motivation:} The goal of this research question is to determine how often human developers explicitly request logging when instructing an AI agent to perform a task. In other words, we quantify the prevalence of explicit logging requirements in the two primary instruction channels that typically guide an agent: the task specification (i.e., the linked issue description) and repository instruction files (e.g., \texttt{AGENTS.md} or \texttt{CLAUDE.md}). This distinction matters because, in traditional development, logging is often governed by implicit norms and learned practices. For example, a junior engineer may naturally add error logs in a \texttt{catch} block without being explicitly instructed to do so. Agents, however, largely rely on their training and what is stated in instructions~(prompts). This difference creates an important ambiguity when interpreting the logging characteristics observed in RQ1. Specifically, it remains unclear whether the observed agentic logging practices are an intrinsic behavior built into the models themselves, or if agents require explicit, lateral logging instructions from developers to implement proper observability. Understanding how frequently humans specify logging, and how agents respond to these instructions, helps disentangle whether logging is a built-in property of the agent or a prompted action. The results of this research question inform whether improving observability in agentic contributions should primarily focus on better agent support for repository-specific logging conventions, on encouraging developers to be more explicit about their logging expectations, or on both.

\begin{figure}[t]
\centering
\includegraphics[width=\textwidth,keepaspectratio,height=5.5cm]{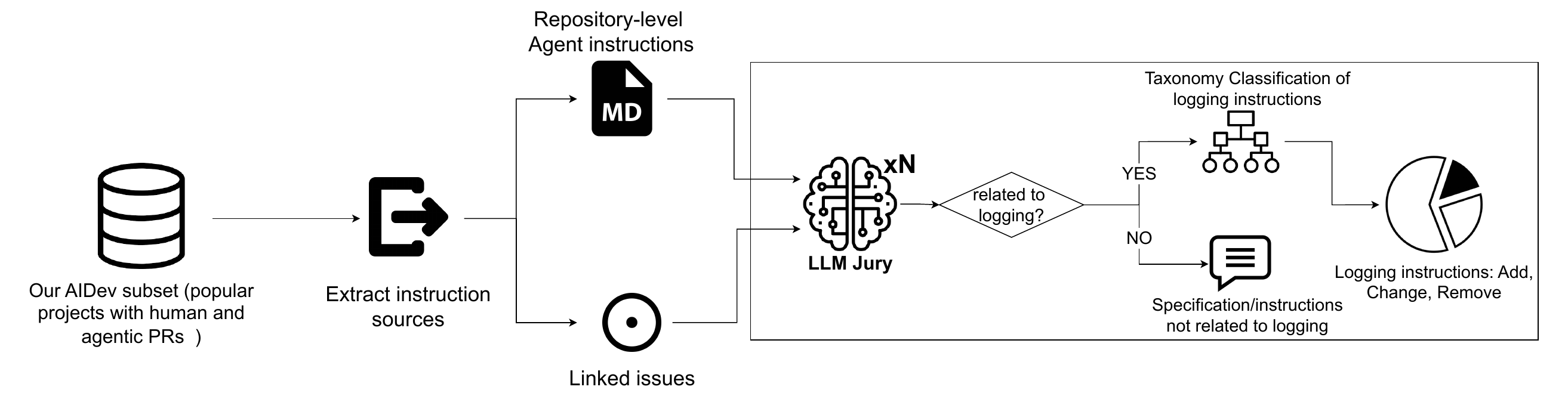}
\caption{Overview of our approach to characterize agentic logging characteristics.}
\label{fig:rq2_meth}
\end{figure}


\smallskip\noindent
\textbf{Approach:} To characterize how humans instruct agents on logging, we analyze two instruction channels: (1) \textit{Task Specifications} from linked issue descriptions (Copilot PRs), and (2) \textit{Repository Instructions} from global instruction files (e.g., \texttt{AGENTS.md}, \texttt{CLAUDE.md}), as illustrated in Figure~\ref{fig:rq2_meth}.  
We compute metrics at two units of analysis: instruction-level and PR-level.

\begin{itemize}
\item \textbf{Instruction-level metrics.} For each detected logging instruction, we measure: (i) \emph{Intent} (\emph{Add}, \emph{Modify}, \emph{Remove}) using the LLM Jury protocol (Section~\ref{sec:llm_jury}), and (ii) \emph{Strength} (\emph{Strong}, \emph{Weak}) via manual labeling by the first two authors (Cohen’s $\kappa=0.96$).

\item \textbf{PR-level metrics.} For each agentic PR, we measure: (i) whether it is \emph{log-instructed} (at least one explicit logging instruction from either Task Specifications or Repository Instructions) or \emph{log-uninstructed}, and (ii) whether the PR changes logging. For log-instructed PRs, we additionally measure \emph{compliance}, i.e., whether the final diff matches the instruction intent (\emph{Add}, \emph{Modify}, \emph{Remove}).
\end{itemize}

These measures capture whether humans provide logging instructions, what behavior they request, and how strong those instructions are. They also allow us to test whether explicit logging instructions are associated with different logging behavior in agentic PRs.

\smallskip\noindent
\textbf{Results:} \textbf{Explicit logging instructions are rare in the two instruction channels we analyze}, as shown in Table~\ref{tab:instruction_breakdown}. Among the 1,308 agentic PRs where at least one instruction channel is observable (a linked issue or a repository instruction file), only 4.7\% (61 out of 1,308) are associated with \textit{any} logging instruction. Furthermore, we observe zero overlap between instruction sources, as 15 PRs receive instructions solely from linked issues, and 46 receive them solely from repository instruction files.
Moreover, we observe that repository instructions often act as cleanup rules, telling agents to delete logs or debug output before finishing. All 10 Remove instructions from repository files originate from a single project (dropseed/plain) and provide the same directive: use statements for debugging, but remove them before committing. This instruction does not ban logging. Instead, it guides the agent to use logs temporarily (for its internal debugging) and then clean them up. Indeed, we observe zero debug statements in the final code. However, we note that this 100\% compliance with removal might be inflated by vacuous compliance, where agents may have simply never added debug statements in the first place, rather than actively removing them.

\begin{table}[t]
\centering
\caption{PR-level breakdown of logging instructions and agent compliance by instruction channel (Task Specifications vs.\ Repository Instructions). Count denotes the number of agentic PRs with at least one explicit logging requirement in that channel.}
\label{tab:instruction_breakdown}
\begin{tabular}{llrrr}
\toprule
\textbf{Instruction channel} & \textbf{Intent} & \textbf{Count} & \textbf{Compliant PRs} & \textbf{Compliance (\%)} \\
\midrule
\textit{Task Specifications} & Add & 5 & 2 & 40.0\% \\
& Modify & 8 & 3 & 37.5\% \\
& Remove & 2 & 0 & 0.0\% \\
\midrule
\textit{Repository Instructions} & Add & 36 & 3 & 8.3\% \\
& Modify & 0 & 0 & 0.0\% \\
& Remove* & 10 & 10 & \textbf{100.0\%} \\
\bottomrule
\multicolumn{5}{l}{\footnotesize{*``Remove debug instrumentation before commit.'' Zero debug instrumentation added = 100\% compliance.}} \\
\end{tabular}
\end{table}

\begin{table}[t] 
\centering 
\caption{Impact of instruction strength on agent compliance (n=15 issue instructions).} 
\label{tab:prompt_quality} 
\small 
\resizebox{\columnwidth}{!}{\begin{tabular}{llrr} 
\toprule 
\textbf{Instruction Strength} & \textbf{Definition} & \textbf{Count} & \textbf{Compliance} \\ 
\midrule 
\textbf{Strong} & Specifics (files, levels, frameworks) & 11 & 27.3\% \\ 
\textbf{Weak} & Generic (``add logs'', ``ensure observability'') & 4 & 50.0\% \\ 
\bottomrule 
\end{tabular}} 
\end{table}

\textbf{Agents show a compliance gap regardless of instruction strength}. In our analysis of task specifications (n=15) and repository instructions (n=46), we find that concrete wording alone does not ensure compliance. As shown in Table~\ref{tab:prompt_quality}, at the issue level, 73.3\% (11 out of 15) of logging instructions are strong, yet compliance among these strong instructions is only 27.3\% (3 out of 11). At the repository-file level, all 46 labeled instructions are strong, but overall compliance remains low (6.5\%, 3 out of 46). Note that while task specifications are Copilot-linked and therefore visible to the model before generation, the visibility of repository instructions may differ across agent workflows. So non-compliance with repository instructions specifically may come from two causes: the file is not surfaced to the agent, or it is surfaced but ignored.

\textbf{Consequently, having a logging instruction in one of the two channels (i.e., task specifications or repository instructions) is not associated with a higher rate of logging changes}. In fact, we find no statistical difference in logging behavior between \textit{log-instructed} and \textit{log-uninstructed} PRs. Agents receiving instructions changed logging in 14.8\% of cases, while uninstructed agents changed logging in 20.8\% of cases, as shown in Table~\ref{tab:governed_prevalence}
. A Pearson's $\chi^2$ test confirms this null result ($\chi^2$=1.32, p=0.25). Thus, simply adding a logging instruction to an issue or repository agent instruction file does not reliably alter the agent's logging behavior.


\begin{table}[t]
\centering
\caption{PR-level logging prevalence by log-instruction status.}
\label{tab:governed_prevalence}
\begin{tabular}{lrrr}
\toprule
\textbf{Group} & \textbf{Logging-change PRs} & \textbf{Total PRs} & \textbf{Prevalence (\%)} \\
\midrule
Log-instructed   & 9   & 61   & 14.8\% \\
Log-uninstructed & 932 & 4489 & 20.8\% \\ 
\midrule
All PRs    & 941 & 4550 & 20.7\% \\
\bottomrule
\end{tabular}
\end{table}

\begin{tcolorbox}[colback=gray!10, colframe=black, title=\textbf{Summary of RQ2}]
Our analysis identifies both a specification gap and a compliance gap in agentic logging. Specifically, 98.7\% of agentic PRs lack logging instructions, and even when explicitly instructed, agents do not comply with requests 67\% of the time. Crucially, this low compliance persists despite 73\% of logging instructions being detailed (specifying log levels, files, or frameworks), these strong instructions yield only 27\% compliance. Practically, this suggests that relying on instruction specificity alone might not ensure observability. Instead, we recommend enforcing logging standards through deterministic CI/CD checks or linters that block uninstrumented code.
\end{tcolorbox}
\subsection*{\textbf{RQ3. \RQIII}}

\textbf{Motivation:}
The goal of this research question is to understand the lifecycle of agentic PRs post-generation. We specifically investigate how logging is regulated during code review and subsequent commits. Unlike functional correctness, which automated tests and CI pipelines can verify, logging quality (e.g., missing context, noise) rarely breaks the build. Consequently, logging issues can easily evade automated scrutiny. Therefore, it remains unclear whether agent-authored logging enters the codebase unexamined, or whether human reviewers and automated bots intervene to request changes and refine logging and observability.

\begin{figure}[t]
\centering

\begin{subfigure}[t]{\textwidth}
    \centering
    \includegraphics[width=\textwidth,keepaspectratio,height=5.5cm]{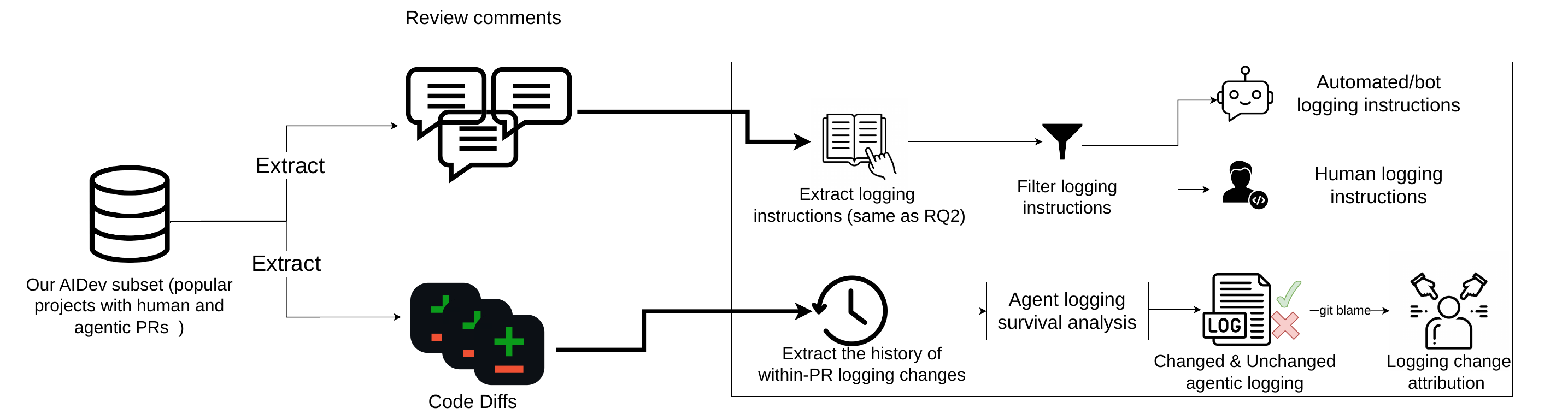}
    \caption{Overview of our approach to tracking post-generation log regulation.}
    \label{fig:rq3_meth}
\end{subfigure}

\vspace{0.6em}

\begin{subfigure}[t]{0.6\textwidth}
    \centering
    \includegraphics[width=\textwidth,keepaspectratio,height=5.5cm]{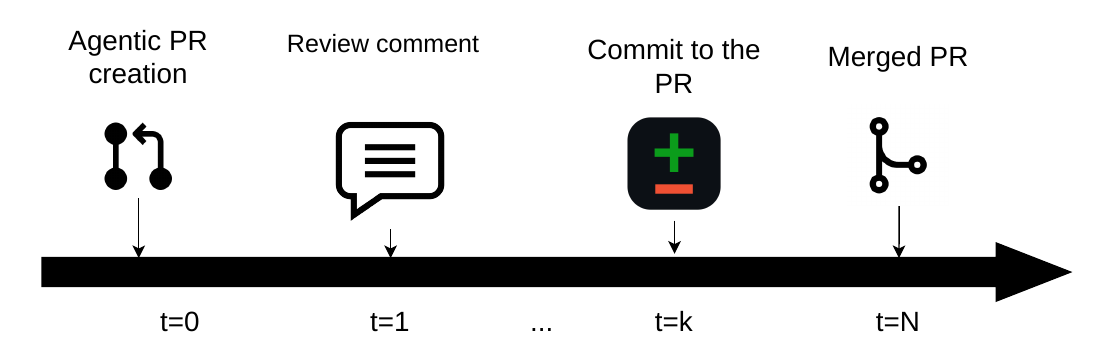}
    \caption{Overview of the PR lifecycle considered for post-generation analysis.}
    \label{fig:lifecycle}
\end{subfigure}

\caption{RQ3 methodology and lifecycle setup.}
\label{fig:rq3_overview}
\end{figure}

\begin{figure}[t]
  \centering
  \begin{subfigure}[t]{0.85\textwidth}
      \centering
      \includegraphics[width=\textwidth,keepaspectratio]{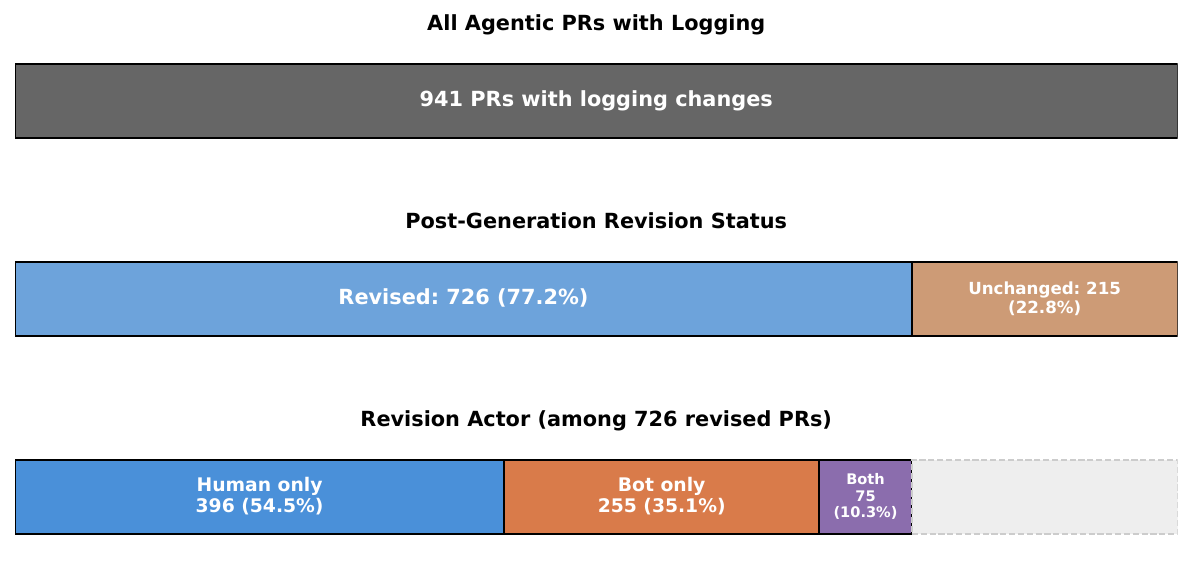}
      \caption{Agentic PRs with logging changes.}
      \label{fig:sankey_agentic}
  \end{subfigure}
  \hfill
  \begin{subfigure}[t]{0.85\textwidth}
      \centering
      \includegraphics[width=\textwidth,keepaspectratio]{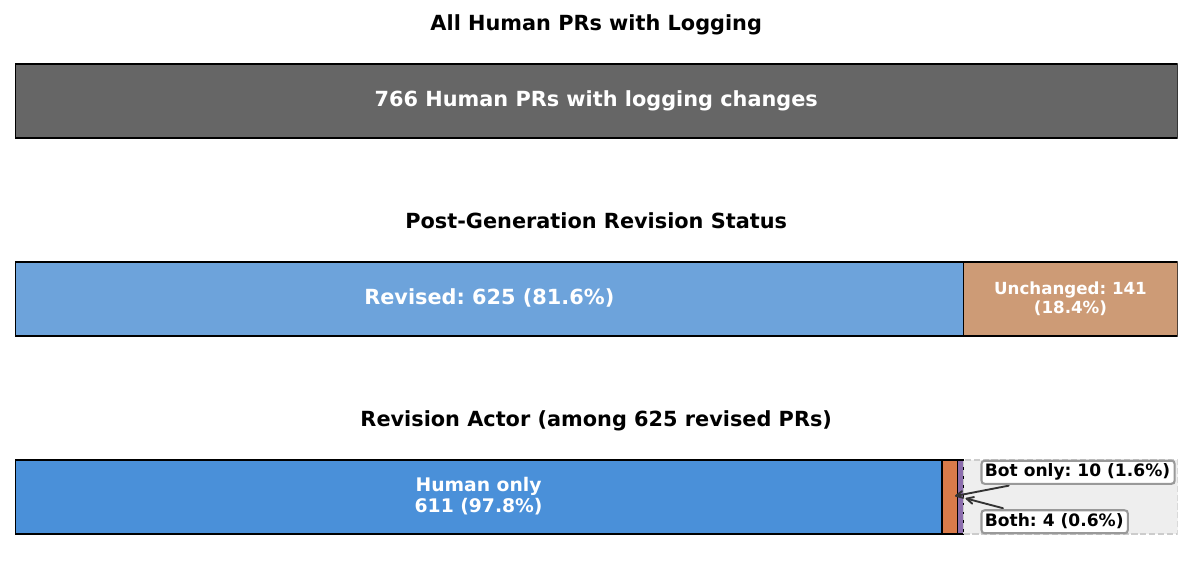}
      \caption{Human PRs with logging changes.}
      \label{fig:sankey_human}
  \end{subfigure}
  \caption{Post-generation logging revision flow for agentic and human PRs.}
  \label{fig:sankey}
  \end{figure}
  
\smallskip\noindent
\textbf{Approach:}
To quantify the post-generation regulation of agentic logging, we analyze the version history and review comments of our dataset, as shown in Figure~\ref{fig:rq3_meth}. Specifically, we perform three primary analyses:

\begin{itemize}
    \item Lifecycle Tracking: We reconstruct the history of every agent-introduced logging statement from its initial commit ($t=0$) to the final merged state, as shown in Figure~\ref{fig:lifecycle}. We attribute every modification or deletion to either a human or a bot using commit authorship metadata. This allows us to quantify how much logging churn is driven by automated iteration versus human intervention.

    \item Logging Instructions Analysis: We analyze review comments across all 4,550 agentic PRs to identify explicit logging feedback and who provides it. A PR is counted as having explicit logging feedback if at least one review comment is labeled as logging-related by our LLM Jury protocol (Section~\ref{sec:llm_jury}). We identify reviewer type from GitHub author metadata (bot account vs. human account). We then classify logging-related comments into \textit{Add} (coverage-seeking), \textit{Modify} (quality-refining), or \textit{Remove} (noise-control). Finally, we report the prevalence of these logging-related comments over both (i) all agentic PRs and (ii) the subset of PRs with logging changes.

    \item Survival Analysis: We use Kaplan-Meier survival analysis~\cite{Kaplan:1958} to study how long agent-generated logs remain unchanged after their introduction. We include only PRs in which the first commit already contains logging changes. The time axis is defined as the number of subsequent commits within the same PR. We mark an event at the first subsequent commit that edits logging (Add, Modify, or Remove). If logging is never edited again, the PR is treated as unchanged up to its last commit. We estimate the survival curve and 95\% confidence intervals using \texttt{scikit-survival}.\footnote{\url{https://scikit-survival.readthedocs.io/}}

\end{itemize}

To contextualize our findings, we apply the exact same three-step approach to our dataset of human-authored PRs. Similar to RQ1, the human PRs considered are from the same repositories and the same time frames as the agentic PRs. This allows us to isolate whether the post-generation regulation and review patterns observed in agentic PRs are unique to AI-generated code, or if they simply reflect the standard review lifecycle of modern software engineering.

\begin{figure}[t]
\centering
\includegraphics[width=0.6\linewidth,keepaspectratio]{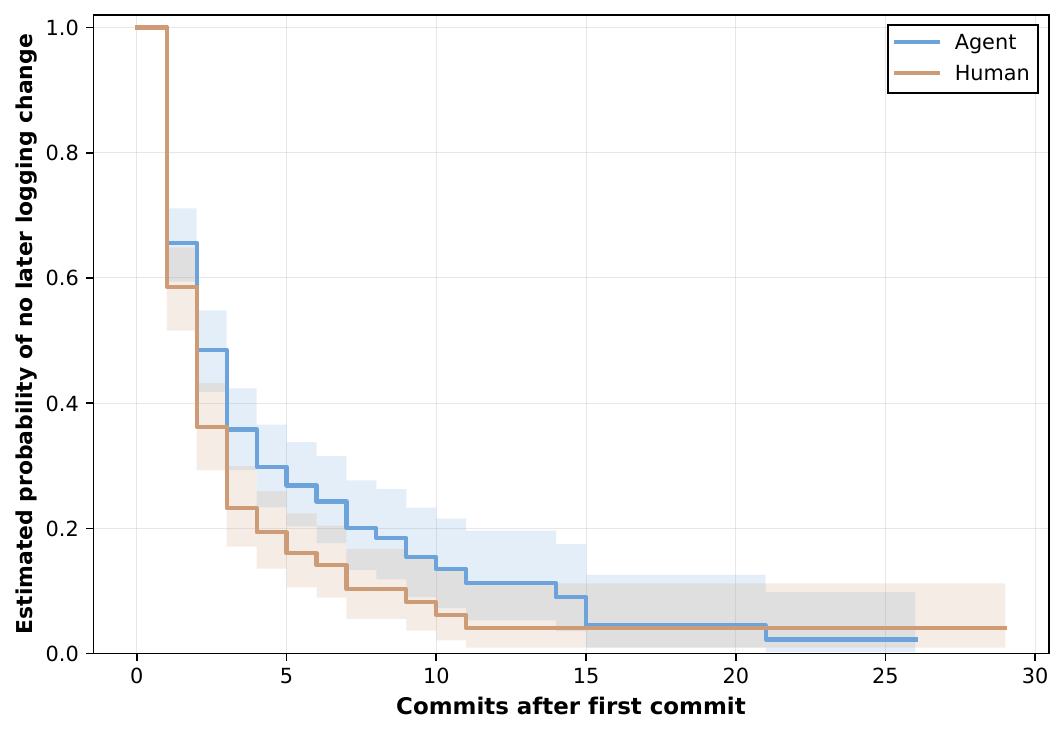}
\caption{Kaplan–Meier survival curves for post-generation logging stability in shared repositories (Agent vs Human). The data includes PRs whose first commit already contains logging changes and an event is the first later commit that modifies logging.}
\label{fig:rq3_log_change_survival}
\end{figure}

\smallskip\noindent
\textbf{Results:} \textbf{Post-generation logging revisions are common in both agentic and human PRs, but the revision actors differ}, as shown in Figure~\ref{fig:sankey}. Among PRs that introduce logging changes, 77.2\% (agentic) and 81.6\% (human-authored) of the PRs are revised in later commits. For the revised PRs, we observe that the actor mix differs sharply. While agentic PRs are mainly split across human-only (54.5\%), bot-only
(35.1\%) revisions, human PRs are almost entirely revised by humans only (97.8\%). At the logging-statement level, humans contribute with 72.5\% post-generation modifications to agentic PRs, versus 99.5\% for human PRs. 

Furthermore, our survival analysis suggests that post-generation logging revisions occur primarily early in the PR lifecycle for both human and agentic PRs, as shown in Figure~\ref{fig:rq3_log_change_survival}. This means that when logging is revised, it is usually adjusted in the first few follow-up commits, while later logging revisions become less common. Notably, we observe a distinct gap in iteration frequency between the two groups (i.e., humans vs. agents). Specifically, the human survival curve drops faster and lower than the agent curve, indicating that human-authored logging undergoes more frequent and rapid post-generation revision. Although agents are still subject to post-generation regulation, their initial logging implementations are more ``sticky'' and less likely to be churned across subsequent commits than those introduced by human developers.

\begin{table}[t]
\centering
\caption{Explicit logging feedback rate in review comments (PR-level).}
\label{tab:rq3_review_feedback_trace}
\begin{tabular}{lrrr}
\toprule
\textbf{PR Group} & \textbf{With Feedback} & \textbf{Total} & \textbf{Rate} \\
\midrule
All agentic PRs & 99 & 4,550 & 2.18\% \\
All human PRs & 71 & 3,276 & 2.17\% \\ \midrule
Agentic PRs with logging changes & 55 & 941 & 5.80\% \\
Human PRs with logging changes & 46 & 766 & 6.00\% \\
\bottomrule
\end{tabular}
\end{table}

\textbf{Explicit logging feedback in review text is rare in both human and agentic PRs}, as shown in Table~\ref{tab:rq3_review_feedback_trace}. We observe no meaningful difference in the prevalence of explicit logging feedback between agentic (2.18\%) and human-authored PRs (2.17\%). Even when restricting to PRs that already contain logging changes in their initial commit, the rates remain low at 5.80\% for agentic PRs and 6.00\% for human PRs. This suggests that logging corrections are usually applied directly in later commits rather than explicitly requested in review text. Furthermore, when logging feedback appears, it is mostly automated and primarily requests modifications. Specifically, bots account for 75.6\% of logging feedback in agentic PRs and 81.1\% in the human cohort. In the agentic subset, human and bot reviewers show similar intent distributions, with \emph{Modify} as the largest category (47.4\% for humans, 44.9\% for bots), followed by \emph{Remove} (28.9\% vs.\ 24.6\%) and \emph{Add} (23.7\% vs.\ 30.5\%). In the human subset, human reviewers also focus on \emph{Modify} (55.0\%), while bot feedback is more skewed toward \emph{Add} (44.2\%).

\begin{figure}[t]
\centering
\includegraphics[width=0.72\textwidth,keepaspectratio]{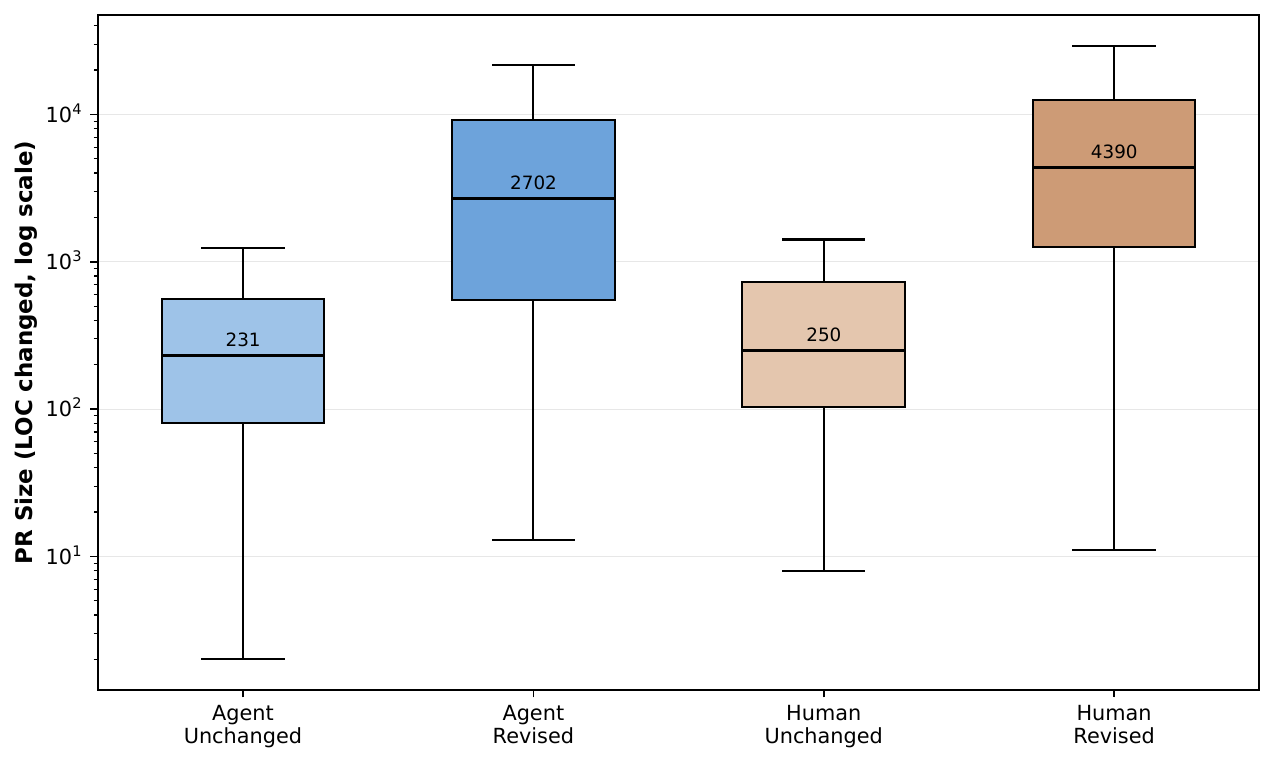}
\caption{PR size versus post-generation logging revisions in agentic and human PRs.}
\label{fig:size_vs_mod}
\end{figure}


\textbf{Logging regulation is mainly concentrated in large PRs}, as shown in Figure~\ref{fig:size_vs_mod}. Specifically, post-generation logging regulation is not evenly distributed, and is concentrated in larger changes. For agentic PRs, those that undergo post-generation log changes are substantially larger (median 2,702 LOC) than those whose logging remains unchanged (median 231 LOC). This difference is statistically significant ($p<0.001$) with a large effect size (Cliff’s $\delta=0.688$). We observe the same pattern for human PRs, as the ones with revised logging have a large effect size of 4,390 LOC, versus 250 LOC for unchanged PRs ($p<0.001$, Cliff’s $\delta=0.726$). We also find a strong positive association between PR size and the amount of post-generation logging modification in both types of PRs (agentic: Spearman $\rho=0.648$, $p<0.001$, human: Spearman $\rho=0.667$, $p<0.001$). Finally, explicit logging feedback is more likely in larger PRs for both groups, and this size effect is stronger for human PRs. In the agentic subset, PRs with feedback have a median size of 320 LOC versus 130 LOC without feedback ($p<0.001$), while the gap is 890.5 LOC versus 114 LOC in the human subset ($p<0.001$).

\begin{tcolorbox}[colback=gray!10, colframe=black, title=\textbf{Summary of RQ3}]
Our analysis reveals that agentic PRs do not significantly reduce the post-generation effort required to regulate logging. Logs introduced by both agents and humans are frequently revised before merging (77.2\% and 81.6\%, respectively). Crucially, the regulation of agentic code remains heavily human-driven as 54.5\% of revised agentic PRs are changed exclusively by humans, who perform 72.5\% of all post-generation log modifications. This persistent reliance on manual intervention demonstrates that current agentic workflows fail to alleviate the human maintenance burden for logging.
\end{tcolorbox}
\section{Implications}
\label{sec:implications}

Our empirical findings reveal a disconnect between how AI coding agents generate code and how software logging observability is maintained. In this section, we discuss the practical implications of these findings for tool builders, software practitioners, and project maintainers.

\subsection{For Tool Builders: Transitioning to Deterministic Guardrails}
Our RQ2 results demonstrate that natural language instruction might be an unreliable mechanism for guiding agentic logging. Developers rarely provide explicit logging instructions (4.7\% of the time). Even when they do, agents fail to comply with constructive requests 67\% of the time. Consequently, tool builders cannot rely solely on prompt engineering or context files (e.g., \texttt{AGENTS.md}) alone to enforce non-functional requirements such as observability. This aligns with the broader findings in the literature demonstrating that the underlying Large Language Models (LLMs) powering these code agents frequently struggle to adhere to strict constraints and complex instructions~\cite{Liu:2023,Zhou:2023}.

To address this, the design of agentic tools should shift from natural language guidance to guardrail-driven development. Tool builders should integrate deterministic enforcement mechanisms directly into the agent's workflow. For example, agents should be required to pass observability-focused static analysis (e.g., linters) or CI/CD checks before submitting a pull request. By treating logging as a hard, verifiable constraint rather than an optional prompt piece, tool builders can ensure that agents produce code that is maintainable by default.

\subsection{For Researchers: Training Agents for Proactive Observability}
Our RQ1 findings show that agents can mimic human error-logging patterns, particularly in exception-handling blocks. However, they significantly underutilize \texttt{INFO}-level logging compared to human developers. This indicates that current models view logging primarily as a reactive mechanism for capturing failures, rather than a proactive tool for tracking normal system states.

This behavioral skew highlights a critical gap in how underlying LLMs are trained or fine-tuned for software engineering tasks. Future research should focus on developing specialized training datasets or reward models that emphasize the semantic value of state-transition logging. For instance, models could be aligned using Reinforcement Learning from Human Feedback (RLHF)~\cite{Ouyang:2022} to capture qualitative developer preferences regarding log clarity, placement, and verbosity. Concurrently, Reinforcement Learning with Verifiable Rewards (RLVR)~\cite{Le:2022} could leverage static analyzers or CI/CD pipelines as objective reward signals to automatically penalize uninstrumented control-flow paths. Ultimately, researchers should train agents not only to write code that passes unit tests, but also to generate the necessary footprints that allow human operators to understand the system's runtime narrative.

\subsection{For Practitioners: Mitigating the Hidden Maintenance Tax}
While agentic coding tools promise increased development velocity, our RQ3 results reveal a hidden maintenance tax. We find that humans perform 72.5\% of post-generation log repairs. Crucially, these interventions are largely implicit, and humans act as ``silent janitors'' who fix logging issues in subsequent commits rather than requesting corrections during code review.

This dynamic is unsustainable for long-term project health. Practitioners and engineering managers should update their code review protocols to explicitly account for agentic contributions. Non-functional requirements, particularly observability, should become first-class items on PR review checklists. Reviewers should be encouraged to reject uninstrumented agentic PRs and prompt the agent to fix the missing logs, rather than silently absorbing the technical debt. Shifting this maintenance burden back to the agent is essential to realizing the full productivity benefits of AI-assisted development.
\section{Threats to Validity}

\subsection{Internal Validity}
Potential subjectivity in classifying the intent of logging instructions poses a threat. To mitigate this risk, we did not rely on a single classifier. Instead, we implemented a rigorous LLM Jury protocol (comprising GPT-4o, GLM-4.7, and DeepSeek) to triangulate the final labels. We iteratively refined our prompting strategy until we achieved substantial agreement (Cohen's $\kappa$ = 0.83) against a manually annotated ground truth, ensuring that our classification is robust and reproducible.

An alternative explanation for non-compliance with repository-level instructions (e.g., \texttt{AGENTS.md}) is that the agent may not have accessed the file due to context window limitations. Our dataset is derived from frontier models (e.g., Claude 3.5 Sonnet, GPT-4o), which feature massive context windows (128k+ tokens) that can easily accommodate typical instruction files. Furthermore, modern agentic scaffolding employs context compaction strategies (e.g., Claude compact). Therefore, the non-compliance we observed is more likely a behavioral alignment issue than a resource limitation.

Ephemeral instructions provided by users before PR generation introduce another threat. We acknowledge a blind spot regarding such instructions delivered via IDE chat interfaces, which are not captured in our dataset. However, we argue that this does not undermine our core finding of a logging gap. If invisible chat instructions were both prevalent and effective, we would expect higher logging prevalence in the final agent-generated code. The fact that agentic PRs still exhibit reduced logging activity suggests that chat-based instructions, if present, fail to drive observability as much as persistent instructions.

\subsection{Construct Validity}
One threat is our reliance on regex-based static analysis to identify logging statements. While pragmatic and well established in logging research, this approach carries the risk of missing dynamic logging patterns or custom wrappers (e.g., a project-specific \texttt{MyLogger.track()} or non-standard libraries). To minimize this threat, we leveraged specialized patterns tailored to the idioms of each target language (Python, Java, and JS/TS) and explicitly excluded build artifacts. We empirically validated the reliability of this approach on a statistically significant sample of 380 code diffs, achieving 96\% precision and 94\% recall.

\subsection{External Validity}
Our study focuses on repositories using Python, Java, and JavaScript/TypeScript. We selected these languages because they have mature tools for studying logging and represent dominant ecosystems in modern software development and AI training datasets~\cite{Octoverse:2024}. Additionally, we restricted our dataset to repositories with at least 100 stars. While our findings may not generalize to small-scale repositories, our conclusions are derived from mature, active projects where logging and observability are typically genuine concerns. Finally, while specific compliance rates may shift as LLM capabilities evolve, the fundamental specification gap we observed, where humans fail to request logging, is a behavioral pattern likely to persist across model generations.
\section{Conclusion}
This paper presents an empirical study of logging practices in agent-generated code, analyzing 4,550 PRs from 81 open-source repositories. We investigate how agents implement logging compared to humans, how they respond to instructions, and how their work is regulated post generation.

We find that agents generally mimic human logging mechanics but exhibit a significant prevalence gap, modifying logs less often than humans in 58.4\% of the studied repositories. Furthermore, natural language instruction is largely ineffective. Explicit logging instructions are rare, appearing in only 4.7\% of PRs, and agents ignore them 67\% of the time. Finally, we identify a hidden maintenance cost, as humans perform 72.5\% of post-generation log repairs, acting as ``silent janitors'' to ensure observability.

These findings suggest that natural language instruction faces a double hurdle: humans rarely provide logging prompts (specification gap), and agents frequently ignore them (compliance gap). Consequently, relying on optional prompts is insufficient to ensure observability. Future work should explore deterministic enforcement mechanisms, such as CI/CD linters, to guarantee that agent-generated code meets production logging standards.

\bibliographystyle{ACM-Reference-Format}
\bibliography{Loganalysis}

\end{document}